\documentclass[floatfix,twocolumn,prl,superscriptaddress]{revtex4}
\usepackage{graphicx,color}
\usepackage{amsmath}
\usepackage{bm}
\usepackage{dcolumn}

\begin{document}

\title{Magnetically Hindered Chain Formation in Transition-Metal Break Junctions}

\author{A.~Thiess}
\email{a.thiess@fz-juelich.de}
\affiliation{Institut f\"ur Festk\"orperforschung and Institute for Advanced Simulation,
Forschungszentrum J\"ulich, D-52425 J\"ulich, Germany}

\author{Y.~Mokrousov}
\affiliation{Institut f\"ur Festk\"orperforschung and Institute for Advanced Simulation,
Forschungszentrum J\"ulich, D-52425 J\"ulich, Germany}

\author{S.~Heinze}
\affiliation{Institut f\"ur Theoretische Physik und Astrophysik, 
Christian-Albrechts-Universit\"at zu Kiel, D-24098 Kiel, Germany}

\author{S.~Bl\"ugel}
\affiliation{Institut f\"ur Festk\"orperforschung and Institute for Advanced Simulation,
Forschungszentrum J\"ulich, D-52425 J\"ulich, Germany}

\date{\today}

\begin{abstract}
Based on first-principles calculations, we demonstrate that magnetism
impedes the formation of long chains in break junctions.
We find a distinct softening of the binding energy of atomic chains 
due to the creation of magnetic moments that
crucially reduces the probability of successful chain formation.
Thereby, we are able to explain the long standing puzzle
why most of the transition-metals do not assemble as long chains in
break junctions and
provide thus an indirect evidence that in general suspended atomic
chains in transition-metal break junctions are magnetic.
\end{abstract}

\maketitle

%===========================================================================
% Motivation (start)
%===========================================================================
One-dimensional systems,
realized experimentally as suspended monatomic chains in break junctions (BJs), 
have altered our conceptional view on atomic scale junctions.
For example, due their enhanced tendency to magnetism they bear 
high potential in the field of spintronics 
by combining the possibility to probe, control \cite{Tosatti:Kondo,Untiedt:Kondo}, and switch the magnetic state 
by spin-polarized electrical currents.
Recent first-principles calculations support these expectations and report on sizeable magnetic 
moments~\cite{Delin:03.1,Mokrousov:06.1,Tung:07.9,Palacios:05,Tosatti-Pt:08} and
giant magneto-crystalline anisotropy energies~\cite{Mokrousov:06.1,Tung:07.9,Tosatti-Pt:08} in suspended 
and free-standing transition-metal (TM) monowires (MWs).
While the formation of long atomic chains of selected TMs and conductance
quantization have been experimentally demonstrated~\cite{Yanson,Smit:01.1,Kizuka}, 
any conclusive evidence of magnetism in chains is still missing.

This lack of evidence seems even more surprising as measurements
on chains deposited on surfaces show univocal signatures of local magnetic moment~\cite{Gambardella:00,Hirjibehedin:06}.
Measuring the magnetoresistance in BJs would serve as a proof for magnetism in atomic-sized contacts. 
In materials, where long MWs can be successfully suspended as reported for Ir and Pt \cite{Smit:01.1,Kizuka:Ir,Shiota:08}, 
the leads are non-magnetic, which prevents the pinning tip-magnetization
and thus the analysis via magnetoresistivity measurements.
On the other hand, so far it has not been shown that breaking contacts of magnetic $3d$-TMs results in 
one-dimensional structures beyond point contacts.

An alternative approach to prove that monatomic chains are magnetic is the search for half-integer 
conductance originating from 100\% spin-polarized conductance channels. Although Rodrigues 
\textit{et al.}~\cite{Rodrigues:02} reported on such a half-integer conductance for Co, Ni and Pt 
BJs, it was shown both experimentally \cite{Untiedt:04} and theoretically \cite{Garcia-Suarez:05} that 
not only magnetic ordering but e.g. also the adsorption of H$_2$ can lead to similar values of conductance. 
In other words, the presence of half-integer conductance is not a unique attribute of full spin-polarization 
and thus cannot serve as a proof of it.

Because of these uncertainties, in this letter we take a new path to address 
the emergence of magnetism in BJs.
We base our study on the most fundamental and easiest accessible experimental quantity: the probability 
for successful chain formation of a given material itself.
The trend arising from numerous BJ experiments is, that monatomic chain formation is most probable 
for late 5$d$ TMs as well as Ag and Au \cite{Kizuka:Ir,Shiota:08,Thijssen:06,Kizuka}. 
To analyze the role
of magnetism for the chain formation in BJs of 3$d$, 4$d$ and 5$d$ TMs
we apply a recently developed material-specific theoretical model~\cite{Nanolett08}
for the formation of long monatomic chains which operates in terms of parameters 
extracted
from \textit{ab initio} calculations.  
By explicitly including and excluding magnetic exchange interactions, 
we prove that magnetism significantly suppresses the chain formation due to 
a substantial reduction of the chain hardness expressed in terms of the 
maximally sustainable break force.
Comparing our results to experimental findings, 
we are able to provide an indirect evidence that chains in BJs are indeed magnetic.

%===========================================================================
% Introduce S and P (start)
%===========================================================================

We will briefly recall the model which allows to investigate the
formation probability of suspended monatomic chains under tension
in BJs~\cite{Nanolett08} before applying it to both magnetic and
non-magnetic chains. 
In this model, chain formation
succeeds if the \textit{criteria for stability} and \textit{
producibility} are met. The \textit{criterion for stability} addresses the rupture of the 
chain via breaking of the bond between two neighboring chain atoms.
The \textit{criterion for producibility} is concerned with the chain 
elongation that is composed of two processes: At first one atom has to be
extracted out of the lead into the chain. This transfer leads to a
reduction of the coordination of this particular atom and consequently
additional external energy is required, which we account for by the
difference $\Delta E_{\mathrm{Lead}}=E_{\mathrm{MW}}-E_{\mathrm{Lead}}$ of the energy of
cohesion for an atom in the lead, $E_{\mathrm{Lead}}$, and in the chain, $E_{\mathrm{MW}}$, 
both at equilibrium distance.
This energy can be stored mechanically and is released in the
second process of relaxing all chain's bonds to a smaller inter-atomic
distance after an additional atom has joined the chain. 
These competing contributions to the total energy of the system determine whether it is
energetically favorable to grow the chain by one atom or not. In order
to apply both criteria only two quantities have
to be known: The binding energy $\mathcal{E}(d)$ of a MW atom as a function of inter-atomic
distance $d$ and the cohesive energy difference $\Delta
E_{\mathrm{Lead}}$. We will show that both parameters depend on the formation 
of magnetic moments.

%===========================================================================
% computational method (start)
%===========================================================================

In order to determine these parameters we carried out spin-polarized and
non-spin-polarized \textit{ab initio} calculations in the generalized gradient 
approximation (GGA) \cite{rpbe} to the density functional theory for selected
$3d$, $4d$, 
and $5d$ TMs, employing the full-potential linearized augmented plane-wave method (FLAPW)
for one-dimensional (1D) systems \cite{Mokrousov:05.1}, as implemented in
the {\tt FLEUR} code \cite{fleur}.
We used the bulk version of this code to calculate the cohesion energy~\cite{cohesion-energy}. Here 
we considered in all cases the true magnetic (3$d$) or non-magnetic (4$d$, 5$d$) bulk ground state 
as the reference configuration. For calculations of the bare MWs in all cases we 
considered the non- (NM), ferro- (FM) and 
antiferromagnetic (AFM) order and included basis functions with
plane waves up to $k_{\mathrm{max}}$ = 4.4~a.u.$^{-1}$ and used 64 $k$ points
in one half of the 1D Brillouin zone. We calculated all $3d$ MWs in the scalar-relativistic approach, while 
spin-orbit coupling was added for all $4d$'s and $5d$'s.

%===========================================================================
% Take a first look at the impact of magnetism on binding energy potentials
% (start)
%===========================================================================

\begin{figure}
\begin{center}
   \includegraphics[angle=0,scale=0.45]{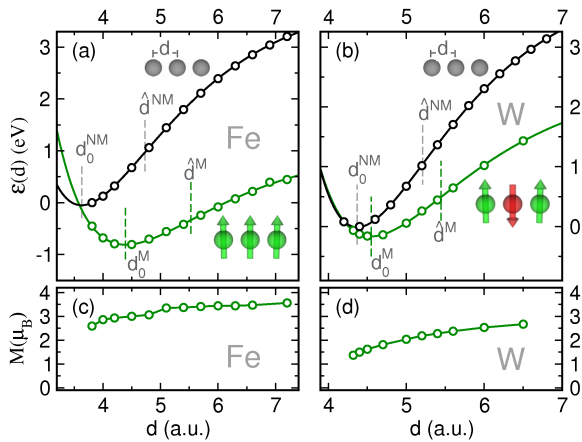}
\end{center}
\caption{\label{FIG1} (color online) Calculated MW energy $\mathcal{E}(d)$ (circles) as a function of 
inter-atomic distance $d$ (throughout the paper given in a.u.=0.0529nm) 
for both non-magnetic (NM) (black) and magnetic (M) state (green) are shown for (a) Fe and (b) W.
The Morse-fit (lines) provides a universal fit to these points well below the required accuracy 
of about 100~meV for chemical bonding and
can be characterized by the equilibrium 
inter-atomic distance $d_0$ and the inflection point $\hat{d}$. $\mathcal{E}(d)$ and the magnetic 
moments $M$ in (c) and (d) correspond to the FM and AFM ground state for 
Fe and W, respectively.}
\end{figure}

The calculated spin moments $M(d)$ and the binding energy $\mathcal{E}(d)$
are shown in Fig.~1 for Fe and W MWs as examples. 
Both chains exhibit sizeable magnetic
moments and the binding energies for the non-magnetic (NM)
and magnetic (M) case differ substantially. 
This is true in particular for 
large distances, 
where the energy difference $\mathcal{E}_\mathrm{sp}(d)=\mathcal{E}_{\mathrm{NM}}(d)
-\mathcal{E}_{\mathrm{M}}(d)$ approaches the
 spin-polarization energy $\mathcal{E}_\mathrm{sp}(\infty)$ of an isolated atom.
A closer look reveals, that 
this overall tendency to magnetism in 1D chains results in finite magnetic moments even in $5d$-TMs already at the NM
equilibrium distance $d_0^{\rm NM}$ or upon small stretching
leading to a magnetic expansion of e.g. $d_0^{\rm M}-d_0^{\rm NM}=0.2$~a.u. for 
W (Fig.~1(a)) and being more pronounced e.g. for Fe (Fig.~1(b)), in agreement with Ref.~\cite{Tung:07.9,Hafner:03.6}. 
Overall, the binding potential energies of M and NM chains differ not only 
by spin-polarization energy and a 
constant shift in $d_0$, but also their slopes are crucially different (Fig.~1(a),(b)).

%===========================================================================
% introduce Morse-potential and explain basic parameters used (start)
%===========================================================================

In order to analyze the binding energy quantitatively, we fit a Morse-potential
\begin{equation} 
\mathcal{E}(d)=\mathcal{E}(\infty)\cdot\left(1-e^{-\gamma(d-d_0)}\right)^2
\end{equation}
to the discrete set of calculated energy points (Fig.~1(a),(b)) for the different magnetic states.
Besides the equilibrium distance $d_0$, the Morse potential can be characterized 
by the following two physically transparent parameters: 
The inflection point $\hat{d}=d_0+\mathrm{ln}2/\gamma$ and 
the break force $F_0=\gamma\cdot\mathcal{E}(\infty)/2$, 
which is the maximal slope $F(\hat{d})$ of the potential. 
Together with $\Delta E_{\mathrm{Lead}}$, which we evaluate 
for a given close-packed surface (Fig.~2(a)), 
these four quantities constitute 
a minimal basis for a realistic description of the chain formation process, providing 
us with an accurate and continuous representation of the binding energy curve.

%===========================================================================
% Discuss impact of magnetism qualitatively based on the parameters shown in 
% Fig 2 (start)
%===========================================================================

\begin{figure}
\begin{center}
   \includegraphics[angle=0,scale=0.42]{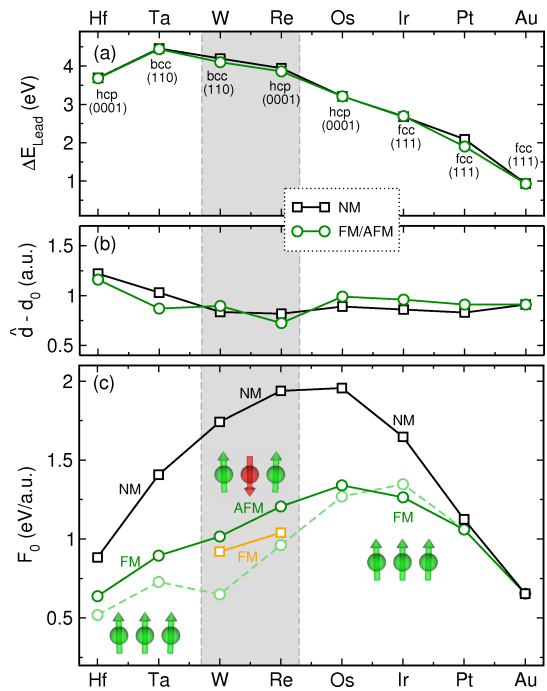}
\end{center}
\caption{\label{FIG2} (color online) (a) Cohesion energy difference $\Delta E_{\mathrm{Lead}}$, (b) difference 
$\hat{d}-d_0=\ln 2/\gamma$ and (c) break force $F_{0}$ for
non-magnetic (marked as NM, black squares) and magnetic (marked as FM or AFM, green circles) 
$5d$-TM chains. 
Broken line in (c) stands for the break force 
$F_0^{\rm{M}}$ calculated according to Eq.~(\ref{equ:F0derived}).
As indicated by the grey shaded area, W and Re MWs reveal an AFM ground state at all inter-atomic 
distances, while the rest of the chains are FM. With orange squares in (c) the break force for 
the FM W and Re MWs is shown for comparison. }
\end{figure}

As we see from Fig.~2, magnetism has a very different impact on these four quantities:
The cohesion energy difference $\Delta E_{\mathrm{Lead}}$ (Fig.~2(a)) of 4$d$ and 5$d$ 
TMs remains almost unchanged for magnetic and non-magnetic chains, 
as these elements are non-magnetic at surfaces and develop only small magnetic moments at $d_0$. 
For 3$d$ TMs the situation is different: 
in the wire the magnetic moments at $d_0$ are already close to the saturated values and larger than for atoms at surfaces, 
which leads to a sizeable gain in spin-polarization energy 
and thereby to a considerable reduction of $\Delta E_{\mathrm{Lead}}$ \cite{Fe-Ru-details}.
The position of $\hat{d}$ with respect to $d_0$ (Fig.~2(b)), 
a quantity crucial for the stability of the wires, reveals only minor 
changes of less than 5\% between calculations with and without spin-polarization.
The crucial impact of magnetism is the softening 
of the binding energy curve, which leads to a significant reduction of the break force $F_0$ 
for all considered 3$d$, 4$d$ (not shown) and 5$d$ TMs (Fig.~2(c)).
For example, the break forces with and without spin-polarization for W  
differ by a factor of two. Both at the beginning and the 
end of the $5d$ series, this ratio reduces according to smaller maximal magnetic moments and is 
equal to one for the non-magnetic noble metal Au.
The break force $F_0$ not only represents the maximal force applicable to a chain, 
but also serves as a measure for the maximal amount of mechanical energy 
which can be gained upon relaxation of its bonds. In any case, high values of $F_0$ favor 
successful chain creation in BJ experiments. 
Thus, as the appearance of magnetism crucially reduces
$F_0$, we can conclude, that the formation of local magnetic moments suppresses chain formation.

%===========================================================================
% answer the question, why F_0 is reduced by magnetism + model (start)
%===========================================================================

In order to capture the origin of the magnetically induced reduction of $F_0$, we
relate the magnetic quantities to the non-magnetic ones.
In the first approximation, the influence of the magnetization on the binding energy potential 
$\mathcal{E}_{\mathrm{sp}}(d)=\mathcal{E}_{\mathrm{NM}}(d)-\mathcal{E}_{\mathrm{M}}(d)$ can be 
attributed to the Hund-type intra-atomic exchange interaction $\frac{1}{2}IM^2(d)$ 
between mostly $d$- but also $s$-electrons on the atomic sites and the inter-atomic Heisenberg 
$J(d)\vec{M}_i(d)\vec{M}_{i+1}(d)$ exchange contributions between the atomic spins.
The binding energy, $\mathcal{E}_{\mathrm{M}}(\infty)$, prefactor in Eq.~(1),  then reads $\mathcal{E}_{\mathrm{M}}(\infty)=
\mathcal{E}_{\mathrm{NM}}(\infty)-
\mathcal{E}_{\mathrm{sp}}(\infty)+
\mathcal{E}_{\mathrm{sp}}(d_0)$, where  
$\mathcal{E}_{\mathrm{sp}}(\infty)=\frac{1}{2}IM^2(\infty)$ is purely given by intra-atomic exchange of a free atom and 
$\mathcal{E}_{\mathrm{sp}}(d_0)=\mathcal{E}_{\mathrm{NM}}(d_0^{\mathrm{NM}})-\mathcal{E}_{\mathrm{M}}(d_0^{\rm{M}})$
is the energy difference of the NM and M states at the corresponding equilibrium distance.
Taking as an example $5d$ TMs we can safely assume even for W, exhibiting the largest equilibrium 
magnetic moment of $M(d_0^{\mathrm{M}})\approx 1\mu_B$ through the series, 
that the impact of magnetism on the equilibrium 
properties of the chains is small and set for simplicity
$d_0^{\mathrm{M}}=d_0^{\mathrm{NM}}$ and $\mathcal{E}_{\mathrm{sp}}(d_0)=0$.
As apparent from~Fig.~2(b), it is also reasonable to assume that $\gamma_{\mathrm{M}}=\gamma_{\mathrm{NM}}$,
in which case the magnetic break force simplifies to:
\begin{equation}
F_0^\mathrm{M}=\frac{\gamma_{\mathrm{M}}}{2}\cdot\mathcal{E}_{\mathrm{M}}(\infty)=
F_0^{\mathrm{NM}}-\frac{\gamma_{\rm{NM}}}{2}\cdot I_d M^2(\infty).
\label{equ:F0derived}
\end{equation}
For $5d$ TM chains we estimated $F_0^\mathrm{M}$ according to Eq.~(\ref{equ:F0derived}) using \textit{ab initio} 
values of the non-magnetic break force $F_0^{\mathrm{NM}}$, magnetic moments at $d=6.5$~a.u.~and the 
atomic exchange integrals between $d$-electrons, $I_d$, from Ref.~[\onlinecite{Brooks}], and plotted it in comparison to the magnetic break 
force determined from \textit{ab initio} in Fig.~2(c).
Good qualitative agreement between the two break force curves underlines that the intra-atomic exchange $I$ is 
the major origin of the magnetically driven reduction of $F_0$, while the inter-atomic exchange plays 
only a minor role. This conclusion is further verified by the observation that changing the
magnetic order from AFM to FM in W and Re MWs results only in a small change of $F_0$ (see Fig.~\ref{FIG2}(c)).

%===========================================================================
% discuss chain formation in (N,d)-phase space
%===========================================================================

We now turn back to the analysis of our \textit{ab initio} results. 
With the knowledge of all key quantities entering the \textit{criteria for stability} and \textit{producibility} 
we can further analyze both criteria in the phase space of the 
number of atoms $N$ and inter-atomic distance $d$ 
presented in Fig.~3. Each of the criteria leads to a distinct region where it is fulfilled and 
accordingly the chain is stable (S) or producible (P). Ideally for a successful chain elongation 
event to happen, both regions (S) and (P) have to overlap (SP).

\begin{figure}
\begin{center}
   \includegraphics[angle=0,scale=0.22]{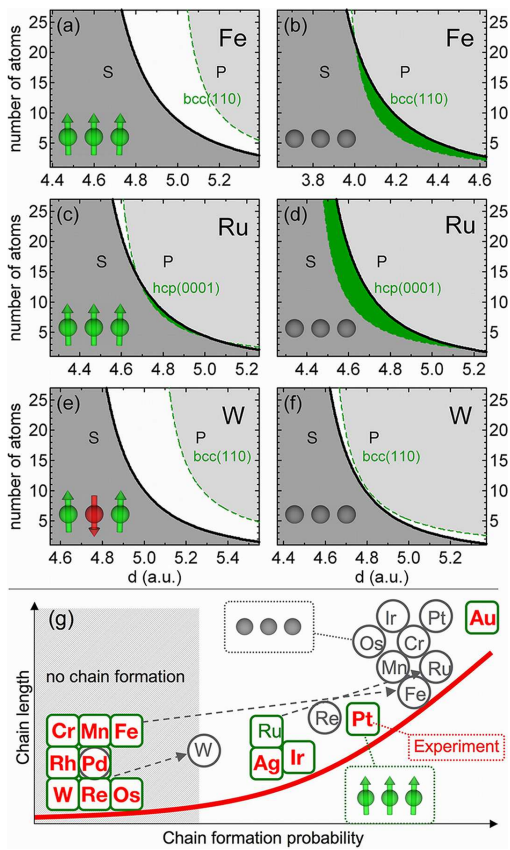}
\end{center}
\caption{\label{FIG3} (color online) Phase diagrams for Fe, Ru, and W-BJs with (a),(c),(e) and
without (b),(d),(f) spin-polarization of the chain atoms.
Plots indicate regions of stability (S, dark gray), producibility
(P, light gray), separated by a white region or overlapping
in the SP region (in green). The starting point along the $x$-axis
$d_0$ and the assumed surface orientations of the leads are indicated. 
The input parameters for W are given in Fig.~2(a)-(c) and 
for Fe and Ru in \cite{Fe-Ru-details}. (g) provides a schematic summary of our results. 
Shown is the chain formation probability, proportional 
to the size of SP-regions, versus chain length, proportional 
to the highest $N$ for which an SP-region exists, with (green) and without 
spin-polarization (gray) in relation to the experimental findings 
(red)~\cite{Untiedt:04,Shiota:08,Thijssen:06,Kizuka:Ir,Kizuka,Landman:90} 
in all cases in arbitrary units relative to Au. 
Arrows indicate (exemplified for Fe, Ru and W) the 
consequence on the chain formation caused by switching off the finite magnetization
in the chain.}
\end{figure}

Comparison of the phase diagrams for different TMs, shown for Fe, Ru, and W in Fig.~3, underlines, that 
the formation of local magnetic moments strongly suppresses the probability of chain 
formation for $3d$, $4d$, and 5$d$ elements.
If we ignore the formation of magnetism 
among the $3d$-, $4d$-, and $5d$-TM series Cr, Mn, Fe, Ru, Rh, Ag, 
Re, Os, Ir, Pt and Au exhibit extensive SP regions (Fig.~3(b),(d)) indicating successful chain formation for these elements.
Even for W (Fig.~3(f)) with bcc(110) electrodes neglecting magnetism results in touching S and P regions, 
indicating chain formation for more open lead structures.
Allowing for the formation of local spin moments the picture changes completely:   
SP regions emerge exclusively for Ru (Fig.~3(c)), Ag, Ir, Pt, and Au while for all other elements  
the S and P regions are clearly separated and no chain formation occurs (Fig.~3(a),(e) and in \cite{Nanolett08}). 
While for Pd and Pt chains the influence of magnetism is small due to relatively small moments
entering Eq.~(\ref{equ:F0derived}), the SP-regions of Ru (Fig.~3(c)) and Ir are considerably 
less extended for magnetic chains than for non-magnetic ones, underlining the suppression of chain formation by magnetism.

%===========================================================================
% proof that chains are magnetic (start)
%===========================================================================

While our predictions based on the assumption that chains in BJs are 
magnetic match and explain the 
experimental findings for successful Ag, Ir, Pt and Au chain formation, 
the results of the model for 
non-magnetic suspended chains contradict the experimental observations at several crucial points (Fig.~3(g)). 
Firstly, nanocontacts of $3d$-TMs such as Fe are reported to form only point-contacts 
with no tendency to form longer chains~\cite{Untiedt:04}. 
Secondly, also BJ-experiments using W as tip-material result only in point-contacts~\cite{Halbritter}, 
moreover, W tips are widely used in STM- and AFM-experiments due to their structural rigidity 
preventing substrate-induced reformations~\cite{Landman:90}.
Thirdly, non-magnetic Ir chains 
would become as long as those of Pt, and almost as long as those of Au, in direct disagreement 
with experiments which report significant decrease of chain formation probability and length 
when going from Au to Ir~\cite{Kizuka,Kizuka:Ir,Shiota:08,Thijssen:06}. These clear contradictions to existing experimental 
evidence lead us to the conclusion that only when chains in BJs are magnetic, the experimentally 
observed trends can be reproduced and explained throughout the $3d$, $4d$, and $5d$ transition- and noble-metal series.
Therefore, by {\it reductio ad contradictum}, comparing theoretical predictions with experimental findings, we provide 
a convincing evidence that TM chains in BJs are magnetic.

\acknowledgments{
We thank J. M. van Ruitenbeek for fruitful discussions.
S.H. thanks the Stifterverband f\"ur die Deutsche Wissenschaft
for financial support.
}


\begin{thebibliography}{99}
\bibitem{cohesion-energy} We calculate $\Delta E_{surf}$ by substracting the surface energy of the corresponding element from Ref.~\cite{Rosengaard:92.9} from the calculated energy difference $\Delta E_{bulk}$ between a bulk atom and a MW atom at $d_0$.
\bibitem{Fe-Ru-details} The resulting set of parameters are for the Fe NM chain: 
$\Delta E_{surf}$=3.07 eV, $F_0$=1.27 eV/a.u., $\hat{d}-d_0^{\rm{NM}}$=1.10 a.u., $d_0^{\rm{NM}}$=3.63 a.u.; 
and for the chain in the magnetic ground state (FM): $\Delta E_{surf}$=2.20 eV, $F_0$=0.58 eV/a.u., 
$\hat{d}-d_0^{\rm{M}}$=1.14 a.u., $d_0^{\rm{M}}$=4.39 a.u.. For the Ru NM chain we obtained: $\Delta E_{surf}$=2.46 eV, 
$F_0$=1.65 eV/a.u., $\hat{d}-d_0^{\rm{NM}}$=0.94 a.u., $d_0^{\rm{NM}}$=4.11 a.u.; and for the chain in the magnetic 
ground state (FM): $\Delta E_{surf}$=2.46 eV, $F_0$=1.02 eV/a.u., $\hat{d}-d_0^{\rm{M}}$=0.99 a.u., $d_0^{\rm{M}}$=4.13 a.u..
\end{thebibliography}
\end{document}